
\font\gross=cmbx10  scaled\magstep2
\font\mittel=cmbx10 scaled\magstep1
\def\gsim{\mathrel{\raise.3ex\hbox{$>$\kern- .75em
                      \lower1ex\hbox{$\sim$}}}}
\def\lsim{\mathrel{\raise.3ex\hbox{$<$\kern-.75em
                      \lower1ex\hbox{$\sim$}}}}

\def\square{\kern 1pt
\vbox{\hrule height 0.6pt\hbox{\vrule width 0.6pt \hskip 3pt
\vbox{\vskip 6pt}\hskip 3pt\vrule width 0.6pt}
\hrule height 0.6pt}\kern 1pt }

\def\sla{\raise.15ex\hbox{$/$}\kern-.72em}


\parskip=\medskipamount
\overfullrule=0pt
\raggedbottom
\def\normalparindent{24pt}
\newif\ifdraft \draftfalse

\nopagenumbers
\footline={\ifnum\pageno=1 {\ifdraft
{\hfil\rm Draft \number\day -\number\month -\number\year}
\else{\hfil}\fi}
\else{\hfil\rm\folio\hfil}\fi}
\def\endpage{\vfill\eject}
\def\beginlinemode{\endmode\begingroup\parskip=0pt
\obeylines\def\\{\par}\def\endmode{\par\endgroup}}
\def\beginparmode{\endmode\begingroup \def\endmode{\par\endgroup}}
\let\endmode=\par
\def\raggedcenter{
              \leftskip=2em plus 6em \rightskip=\leftskip
             \parindent=0pt \parfillskip=0pt \spaceskip=.3333em
            \xspaceskip=.5em\pretolerance=9999 \tolerance=9999
              \hyphenpenalty=9999 \exhyphenpenalty=9999 }
\def\\{\cr}
\let\rawfootnote=\footnote
\def\footnote#1#2{{\parindent=0pt\parskip=0pt
\rawfootnote{#1}{#2\hfill\vrule height 0pt depth 6pt width 0pt}}}
\def\title
{\null\vskip 3pt plus 0.2fill\beginlinemode\raggedcenter\gross}
\def\author{\vskip 3pt plus 0.2fill \beginlinemode\raggedcenter}

\def\abstract{\vskip 3pt plus 0.3fill \beginparmode{\noindent
{\mittel Abstract}:~}  }
\def\endtitlepage{\endpage\body}
\def\body{\beginparmode\parindent=\normalparindent}
\def\head#1{\par\goodbreak{\immediate\write16{#1}
\vskip 0.4cm{\noindent\gross
#1}\par}\nobreak\nobreak\nobreak\nobreak}

\def\finalcite{\citeall\ref\citeall\Ref}
\newif\ifannpstyle
\newif\ifprdstyle
\newif\ifplbstyle
\newif\ifwsstyle

\gdef\refto#1{\ifprdstyle  $^{\[#1] }$ \else
              \ifwsstyle$^{\[#1]}$  \else
              \ifannpstyle $~[\[#1] ]$ \else
              \ifplbstyle  $~[\[#1] ]$ \else
                                         $^{[\[#1] ]}$\fi\fi\fi\fi}
\gdef\refis#1{\ifprdstyle \item{~$^{#1}$}\else
              \ifwsstyle \item{#1.} \else
              \ifplbstyle\item{~[#1]} \else
              \ifannpstyle \item{#1.} \else
                              \item{#1.\ }\fi\fi\fi\fi }
\gdef\journal#1,#2,#3,#4.{
           \ifprdstyle {#1~}{\bf #2}, #3 (#4).\else
           \ifwsstyle {\it #1~}{\bf #2~} (#4) #3.\else
           \ifplbstyle {#1~}{#2~} (#4) #3.\else
           \ifannpstyle {\sl #1~}{\bf #2~} (#4), #3.\else
                       {\sl #1~}{\bf #2}, #3 (#4)\fi\fi\fi\fi}

\def\ref#1{Ref.~#1}
\def\Ref#1{Ref.~#1}
\def\cite#1{{#1}}\def\[#1]{\cite{#1}}

\def\eq#1{Eq.~\(#1)}
\def\Eq#1{Eq.~\(#1)}
\def\(#1){(\call{#1})}
\def\call#1{{#1}}\def\taghead#1{{#1}}

\def\references{\head{References}
\beginparmode\frenchspacing\parskip=0pt}
\def\endreferences{\body}
\def\endit{\endmode\vfill\supereject}\let\endpaper=\endit
\def\pr{\journal Phys. Rev.,}
\def\prd{\journal Phys. Rev. D,}
\def\prl{\journal Phys. Rev. Lett.,}

\def\npb{\journal Nucl. Phys. B,}
\def\ijmpa{\journal Int. J. Mod. Phys. A,}

\def\plb{\journal Phys. Lett. B,}


\def\kndir{Universit\"at Konstanz, Fakult\"at f\"ur Physik, Postfach
5560,
D-7750, Konstanz, Germany.\\ e-mail: phlousto@dknkurz1.}

\def\meudir{Observatoire de Paris, Section Meudon, Demirm, UA 336
Laboratoire
associ\' e au CNRS,\\ Observatoire de Meudon et \' Ecole Normale
Sup\' erieure, 92195 Meudon Principal Cedex, FRANCE.}

\def\iafe{Permanent Address: IAFE, Cas. Corr. 67, Suc. 28, 1428
Buenos Aires,
ARGENTINA.\\ e-mail: lousto@iafe.edu.ar.}

\def\mangos{This work was partially supported
by the Directorate General for
Science, Research and Development of
the Commission of the European Communities
and by the Alexander von Humboldt Foundation.}

\catcode`@=11
\newcount\r@fcount \r@fcount=0\newcount\r@fcurr
\immediate\newwrite\reffile\newif\ifr@ffile\r@ffilefalse
\def\w@rnwrite#1{\ifr@ffile\immediate\write\reffile{#1}\fi\message{#1
}}
\def\writer@f#1>>{}
\def\referencefile{\r@ffiletrue\immediate\openout\reffile=\jobname.re
f%
  \def\writer@f##1>>{\ifr@ffile\immediate\write\reffile%
    {\noexpand\refis{##1} = \csname r@fnum##1\endcsname = %
     \expandafter\expandafter\expandafter\strip@t\expandafter%
\meaning\csname r@ftext\csname r@fnum##1\endcsname\endcsname}\fi}%
  \def\strip@t##1>>{}}

\def\citeall#1{\xdef#1##1{#1{\noexpand\cite{##1}}}}
\def\cite#1{\each@rg\citer@nge{#1}}
\def\each@rg#1#2{{\let\thecsname=#1\expandafter\first@rg#2,\end,}}
\def\first@rg#1,{\thecsname{#1}\apply@rg}
\def\apply@rg#1,{\ifx\end#1\let\next=\relax%
\else,\thecsname{#1}\let\next=\apply@rg\fi\next}%
\def\citer@nge#1{\citedor@nge#1-\end-}
\def\citer@ngeat#1\end-{#1}
\def\citedor@nge#1-#2-{\ifx\end#2\r@featspace#1
  \else\citel@@p{#1}{#2}\citer@ngeat\fi}
\def\citel@@p#1#2{\ifnum#1>#2{\errmessage
{Reference range #1-#2\space is bad.}
\errhelp{If you cite a series of references
by the notation M-N, then M and
N must be integers, and N must be greater than or equal to M.}}\else%
{\count0=#1\count1=#2\advance\count1
by1\relax\expandafter\r@fcite\the\count0,%
  \loop\advance\count0 by1\relax
    \ifnum\count0<\count1,\expandafter\r@fcite\the\count0,%
  \repeat}\fi}
\def\r@featspace#1#2 {\r@fcite#1#2,}
\def\r@fcite#1,{\ifuncit@d{#1}
    \expandafter\gdef\csname r@ftext\number\r@fcount\endcsname%
{\message{Reference #1 to be supplied.}
\writer@f#1>>#1 to be supplied.\par
     }\fi\csname r@fnum#1\endcsname}
\def\ifuncit@d#1{\expandafter\ifx\csname r@fnum#1\endcsname\relax%
\global\advance\r@fcount by1%
\expandafter\xdef\csname r@fnum#1\endcsname{\number\r@fcount}}
\let\r@fis=\refis   \def\refis#1#2#3\par{\ifuncit@d{#1}%
\w@rnwrite{Reference #1=\number\r@fcount
\space is not cited up to now.}\fi%
  \expandafter\gdef\csname r@ftext\csname
  r@fnum#1\endcsname\endcsname%
  {\writer@f#1>>#2#3\par}}
\def\r@ferr{\endreferences\errmessage{I was expecting to see
\noexpand\endreferences before now;  I have inserted it here.}}
\let\r@ferences=\references
\def\references{\r@ferences\def\endmode{\r@ferr\par\endgroup}}
\let\endr@ferences=\endreferences
\def\endreferences{\r@fcurr=0{\loop\ifnum\r@fcurr<\r@fcount
\advance\r@fcurr by 1\relax\expandafter\r@fis
\expandafter{\number\r@fcurr}%
    \csname r@ftext\number\r@fcurr\endcsname%
  \repeat}\gdef\r@ferr{}\endr@ferences}
\let\r@fend=\endpaper\gdef\endpaper{\ifr@ffile
\immediate\write16{Cross References written on
[]\jobname.REF.}\fi\r@fend}
\catcode`@=12
\finalcite
\catcode`@=11
\newcount\tagnumber\tagnumber=0
\immediate\newwrite\eqnfile\newif\if@qnfile\@qnfilefalse
\def\write@qn#1{}\def\writenew@qn#1{}
\def\w@rnwrite#1{\write@qn{#1}\message{#1}}
\def\@rrwrite#1{\write@qn{#1}\errmessage{#1}}
\def\taghead#1{\gdef\t@ghead{#1}\global\tagnumber=0}
\def\t@ghead{}\expandafter\def\csname @qnnum-3\endcsname
  {{\t@ghead\advance\tagnumber by -3\relax\number\tagnumber}}
\expandafter\def\csname @qnnum-2\endcsname
  {{\t@ghead\advance\tagnumber by -2\relax\number\tagnumber}}
\expandafter\def\csname @qnnum-1\endcsname
  {{\t@ghead\advance\tagnumber by -1\relax\number\tagnumber}}
\expandafter\def\csname @qnnum0\endcsname
  {\t@ghead\number\tagnumber}
\expandafter\def\csname @qnnum+1\endcsname
  {{\t@ghead\advance\tagnumber by 1\relax\number\tagnumber}}
\expandafter\def\csname @qnnum+2\endcsname
  {{\t@ghead\advance\tagnumber by 2\relax\number\tagnumber}}
\expandafter\def\csname @qnnum+3\endcsname
  {{\t@ghead\advance\tagnumber by 3\relax\number\tagnumber}}
\def\equationfile{\@qnfiletrue\immediate
\openout\eqnfile=\jobname.eqn%
  \def\write@qn##1{\if@qnfile\immediate\write\eqnfile{##1}\fi}
  \def\writenew@qn##1{\if@qnfile\immediate\write\eqnfile
    {\noexpand\tag{##1} = (\t@ghead\number\tagnumber)}\fi}}
\def\callall#1{\xdef#1##1{#1{\noexpand\call{##1}}}}
\def\call#1{\each@rg\callr@nge{#1}}
\def\each@rg#1#2{{\let\thecsname=#1\expandafter\first@rg#2,\end,}}
\def\first@rg#1,{\thecsname{#1}\apply@rg}
\def\apply@rg#1,{\ifx\end#1\let\next=\relax%
\else,\thecsname{#1}\let\next=\apply@rg\fi\next}
\def\callr@nge#1{\calldor@nge#1-\end-}\def\callr@ngeat#1\end-{#1}
\def\calldor@nge#1-#2-{\ifx\end#2\@qneatspace#1 %
  \else\calll@@p{#1}{#2}\callr@ngeat\fi}
\def\calll@@p#1#2{\ifnum#1>#2{\@rrwrite{Equation range #1-#2\space is
bad.}
\errhelp{If you call a series of equations by the notation M-N, then
M and
N must be integers, and N must be greater than or equal to M.}}\else%
{\count0=#1\count1=#2\advance\count1 by1
\relax\expandafter\@qncall\the\count0,%
  \loop\advance\count0 by1\relax%
 \ifnum\count0<\count1,\expandafter\@qncall\the\count0,  \repeat}\fi}
\def\@qneatspace#1#2 {\@qncall#1#2,}
\def\@qncall#1,{\ifunc@lled{#1}{\def\next{#1}\ifx\next\empty\else
\w@rnwrite{Equation number \noexpand\(>>#1<<) has not been defined
yet.}
 >>#1<<\fi}\else\csname @qnnum#1\endcsname\fi}
\let\eqnono=\eqno\def\eqno(#1){\tag#1}
\def\tag#1$${\eqnono(\displayt@g#1 )$$}
\def\aligntag#1\endaligntag  $$
{\gdef\tag##1\\{&(##1 )\cr}\eqalignno{#1\\}$$
  \gdef\tag##1$${\eqnono(\displayt@g##1 )$$}}
  
\def\eqalignno#1{\displ@y \tabskip\centering
  \halign to\displaywidth{\hfil$\displaystyle{##}$\tabskip\z@skip
    &$\displaystyle{{}##}$\hfil\tabskip\centering
    &\llap{$\displayt@gpar##$}\tabskip\z@skip\crcr
    #1\crcr}}
\def\displayt@gpar(#1){(\displayt@g#1 )}
\def\displayt@g#1 {\rm\ifunc@lled{#1}\global\advance\tagnumber by1
        {\def\next{#1}\ifx\next\empty\else\expandafter
   \xdef\csname @qnnum#1\endcsname{\t@ghead\number\tagnumber}\fi}%
  \writenew@qn{#1}\t@ghead\number\tagnumber\else
        {\edef\next{\t@ghead\number\tagnumber}%
        \expandafter\ifx\csname @qnnum#1\endcsname\next\else
 \w@rnwrite{Equation \noexpand\tag{#1} is a duplicate number.}\fi}%
  \csname @qnnum#1\endcsname\fi}
\def\eqnoa(#1){\global\advance\tagnumber by1\multitag{#1}{a}}
\def\eqnob(#1){\multitag{#1}{b}}
\def\eqnoc(#1){\multitag{#1}{c}}
\def\eqnod(#1){\multitag{#1}{d}}
\def\multitag#1#2$${\eqnono(\multidisplayt@g{#1}{#2} )$$}
\def\multidisplayt@g#1#2 {\rm\ifunc@lled{#1}
        {\def\next{#1}\ifx\next\empty\else\expandafter
    \xdef\csname @qnnum#1\endcsname{\t@ghead\number\tagnumber b}\fi}%
  \writenew@qn{#1}\t@ghead\number\tagnumber #2\else
        {\edef\next{\t@ghead\number\tagnumber #2}%
 \expandafter\ifx\csname @qnnum#1\endcsname\next\else
\w@rnwrite{Equation \noexpand\multitag{#1}{#2} is a duplicate
number.}
\fi}%
  \csname @qnnum#1\endcsname\fi}
\def\ifunc@lled#1{\expandafter\ifx\csname @qnnum#1\endcsname\relax}
\let\@qnend=\end\gdef\end{\if@qnfile
\immediate\write16{Equation numbers written on []\jobname.EQN.}
\fi\@qnend}

\magnification=1200
\baselineskip=18pt
\title
{THE QUANTIZATION OF THE SPACETIME GEOMETRY
GENERATED BY PLANCKIAN ENERGY PARTICLES}
\bigskip\bigskip
\author
C.O.LOUST\'O\footnote{$^{a}$}{\kndir}{$^,$}\footnote{$^{b}$}{\iafe}
and

N.S\'ANCHEZ\footnote{$^c$}{\meudir}
\bigskip\bigskip
\abstract
We study the quantization of the curved spacetime created by
ultrarelativistic particles at Planckian energies. We consider
a minisuperspace model based on the classical shock wave metric
generated by these particles, and for which
the Wheeler - De Witt equation is solved exactly. The wave function
of the geometry is a Bessel function whose argument is the classical
action. This allows us to describe not only the semiclassical regime
$(S\to\infty),$ but also the strong quantum regime $(S\to0).$
We analyze the
interaction with a scalar field $\phi$ and apply the third
quantization formalism to it. The quantum
gravity effects make the system to evolve from a highly
curved semiclassical
geometry (a gravitational wave metric) into a strongly
quantum state represented by
a weakly curved geometry (essentially flat spacetime).

\endtitlepage
\baselineskip=18pt
\head{1. Introduction}

The spacetime geometry created by ultrarelativistic sources, that is,
the
gravitational shock wave spacetimes, have arisen great
interest recently\refto{1,2,3,4,5,dv,6,dvs,7,np1,8,9}.
These backgrounds are relevant to
describe the particle scattering at the Planck energy scale. Quantum
scattering of particle fields and strings by this class of metrics
have been studied in the approximation in which the geometry is
treated
classically, i.e., as a background field.

A further step in this direction is to quantize the geometry itself.
As it is known, so far there is not a quantum theory of gravitation
to fully carry out this program. Although a
conventional quantum field theory of gravitation lacks to be
renormalizable,
information about the quantization of the spacetime geometry can be
obtained
by solving the Wheeler - De Witt equation\refto{wdw}:
$$
\left[G_{ijkl}{\delta\over \delta g_{ij}}{\delta\over \delta_{kl}}+
\sqrt{^{(3)}g}~^{(3)}R\right]\Psi(^{(3)} {\cal G})=0~. \eqno(1)
$$
In the canonical description of General Relativity,
the space-time metric has the $3+1$ decomposition
$$
d^2s=(N^2-N_iN^i)dt^2-2N_idx^idt+g_{ij}dx^idx^j~, \eqno(2)
$$
that is
$$
g_{\mu\nu}=\pmatrix{-N^{-2} & N^{-2}\cr
                    N^{-2}N^i & g^{ij}-N^{-2}N_iN_j\cr}~.    \eqno(3)
$$
In the region between two space like hypersurfaces $t=t_i$ and
$t=t_f$,
the Einstein equations determine the sequence of three-geometries
$^{(3)} {\cal G}$ on the space-like
surfaces of constant $t$, the dynamical important
object being the 3-geometry $^{(3)} {\cal G}$. The dynamics of the
gravitational
field is entirely described by the so called Hamiltonian constraint
${\cal H}=0$.
In the quantum theory, this becomes an equation for the state vector
${\cal H}\Psi=0$, which takes the form of the functional
differential
equation \(1). There are also the others constraints of the classical
theory, but at the quantum level, they express merely the gauge
invariance
on $\Psi$. Classically, one can know both $^{(3)} {\cal G}(t_0)$ and
${\partial\over \partial t}^{(3)} {\cal G}(t_0)$
at some time parameter $t_0$ and
determine the 4-geometry $^{(4)}{\cal G}$, but quantum mechanically,
one can
only know $^{(3)} {\cal G}(t_0)$ {\it or}
${\partial \over \partial t}^{(3)} {\cal G}(t_0)$
and therefore, one has a certain probability
for $^{(3)} {\cal G}(t)$. The manifold of all
possible $^{(3)} {\cal G}$ - the so called superspace - in which each
point is a
metric $g_{ij}(\vec x)$, has the metric
$$
G_{ijkl}=\sqrt{g}(g_{ik}g_{jl}+g_{il}g_{jk}-g_{ij}g_{kl})~, \eqno(3')
$$
with signature $(-,~+,~+,~+,~+,~+)$. In order to solve \eq{1}, one
considers
spacetime symmetries and restricts the degrees of freedom to be
quantized
(minisuperspace models).

\head{2. Pure Gravity Model}

Let us quantize now the gravitational
shock wave geometries. The minisuperspace metric to quantize is
$$
ds^2=-(N^2-N_uN^u)dv^2+2N_ududv+F(\rho,u)du^2+dx^2+dy^2 \eqno(4)
$$
where $ u=z-t$, $v=z+t$ are null variables and $\rho=\sqrt{x^2+y^2}$.
The classical shock wave metrics have the form of \Eq{4}
with\refto{np1}
$$
F(u,\rho)=f(\rho)\delta(u)~,  \eqno(5)
$$
and the Lagrange multipliers taken $N_u=1$ , $N=0$.
This expression of the metric is generic for any ultrarelativistic
source; its
particularities entering only in the form of the function $f(\rho)$.
This
expression can be extended to D - dimensions and to shock waves
superimposed to curved
backgrounds\refto{4,6}. \Eq{4} can also represent a sourceless plane
gravitational wave. In such case
$$F(x,y,u)=(x^2-y^2)D(u)~,$$
$D(u)$ being an arbitrary function of $u$.

The metric on superspace has the components:
$$
\eqalign{G_{1111}&={1\over 2}F^{3/2}~,\cr
G_{1122}&=G_{1133}=-{1\over 2}F^{1/2}~,\cr
G_{2222}&=G_{3333}=-G_{2233}={1\over 2}F^{-1/2}~,\cr}
$$
the others components vanish.

The three - curvature $~^{(3)}R$ is given by
$$~^{(3)}R=-g^{11}\nabla^2_{\perp}g_{11}+{1\over 2}(g^{11})^{-2}(\vec
\nabla_{\perp}g_{11})^2=
{1\over F}[-\nabla^2_{\perp}F+{1\over 2F}(\vec\nabla_{\perp}F)^2]
$$
where the subscript $\perp$ refers to the (2, 3) spatial transverse
coordinates.

Thus, the \Eq{1} in our case reads,
$$
\left[-{1\over2}F^{3/2}{\delta^2\over\delta F^2}+F^{1/2}\left(F^{-1}
\nabla_{\perp}^2F-{F^{-2}\over2}\left(\vec\nabla_{\perp}
F\right) ^2\right)\right]\Psi(F)=0~.\eqno(12)
$$
Here, for the sake of convenience, we have taken a 3+1 decomposition
with respect to the hypersurface
$v=const$, that will play the role of time.
Notice also that the supermomentum constraint is already
satisfied classically, since for the metric \(4), one has $R_{vu}=0$.
Thus,
we do not need to consider the quantization of this
equation\refto{ML}.

Let us consider now a minisuperspace model by freezing out all the
transversal degrees of freedom and quantizing the longitudinal one,
i.e.
we write,
$$
F(u,\rho)=f(\rho)D(u)~,\eqno(13)
$$
where $f(\rho)$ is the profile function characterizing the classical
shock wave metric\refto{np1}. The function $D(u)$ will represent
the degree of freedom to quantize (classically,
$D(u)\equiv\delta(u)$).

By discretizing the variable $u\to u_n$ and making the
change of variable $s^2(u_n)=D(u_n)$ we can transform the functional
differential equation \(12) into the following ordinary differential
equation,
$$
\left[{d^2\over ds^2}+(2p-1)s^{-1}{d\over ds}+4C(\rho)\right]
\Psi(s_n)=0~,\eqno(18)
$$
where $p$ accounts for the arbitrariness in the operator ordering,
and
$$
C(\rho)\doteq-2\left(
\nabla_{\perp}^2f-{f^{-1}\over2}\left(\vec\nabla_{\perp}
f\right) ^2\right)=2f~^{(3)}R(\rho)~.\eqno(16)
$$
The whole wave function will read
$$
\Psi(D)=\prod_n\Psi(D(u_n))~.
$$
For $p=1$ (which corresponds to the Laplacian ordering prescription),
\Eq{18}
can be brought into a Bessel equation of index $\nu=0$.
We choose the solution that remains finite when $D\to0$. Thus, our
solution reads,
$$
\Psi(F_n)=J_0\left(2\sqrt{C(\rho)D(u_n)}\right)=J_0\left(2\sqrt{
2~^{(3)}RF(\rho,u_n)}\right)~.\eqno(19)
$$

We can obtain the semiclassical limit by considering large arguments
of the Bessel function in equation \(19). Thus,
$$
\Psi(F)\simeq\left(\pi\sqrt{2~^{(3)}RF(\rho,u_n)}\right)^{-1/2}
\cos\left[2\sqrt{2~^{(3)}RF(\rho,u_n)}-\pi/4\right]~.\eqno(21)
$$

On the other hand, the semiclassical regime can also be directly
studied from the Hamilton - Jacobi equation for our system,
$$
-{1\over2}F^{3/2}\left({dS\over
dF}\right)^2+F^{1/2}{}~^{(3)}R(\rho)=0~,
$$
where $S$ is the Hamilton - Jacobi principal function that by direct
integration reads,
$$
S=\pm2\sqrt{2~^{(3)}R F}~.\eqno(23)
$$

We see that the argument of the wave function \(19) is the {\it
action}
of the classical Hamilton - Jacobi equation. Thus, for large $S$,
the Bessel function has an oscillatory behavior and the wave function
\Eq{19}, is given by
$$
\Psi\simeq{A\over{S}}\exp\{iS/\hbar\}+
{B\over{S}}\exp\{-iS/\hbar\}~.\eqno(24)
$$
that is, \Eq{21} gives the right semiclassical limit.

Semiclassically, the wave function $\Psi$ is picked around the
classical
metric geometry. It appears natural to interpret the oscillatory
behavior
\Eq{24} as describing $\Psi$ in the classical allowed regime.
The classically forbidden  regime, instead, appears with a {\it real}
exponential behavior (Euclidean signature region).

For small $S$, $(S\ll \hbar)$, the behavior is non - oscillatory. In
particular, the Bessel function reaches a finite value in the limit
$S\to0$, i.e. $J_0(0)=1$. This can be interpreted as a genuine
quantum behavior, as opposite to the large $S$ sector which describes
the semiclassical regime. Notice that $S\to0$, i.e. $D(u)=0$, is
flat spacetime and that this appears in the strong quantum regime.
For $F>0$ the action is real and \Eq{24} describes
$\Psi$ in the classically allowed regime. For each $S$, i.e. each
$F$, $\Psi$ describes a classical configuration. All configurations
are classically allowed.

There, $\mid\Psi\mid^2$ (which to some extent can be
seen as proportional to a probability), is a maximum. We will come
back
to this interpretation by the end of the paper.

For $~^{(3)}R>0$ the action is real and \Eq{21} describes the
classically
allowed regime. If $~^{(3)}R<0$ then
$\Psi(S\to\infty)\sim\exp\{S\}/{S}$. This is an exponential growing
of the semiclassical wave function. It is the opposite of what
happens
in the case of tunnel effect, and like the situation of the falling
of
a particle into a well potential, indicating the presence of an
instability. That is, the classically forbidden regime does not
corresponds to tunnel effect, but to an unstable (exponentially
growing)
behavior.

\head{3. Gravity plus matter}

We will include a scalar field in the analysis of the wave function.
The Wheeler - De Witt equation in this case reads
$$
\left\{-{\delta^2\over\delta F^2}-{2~^{(3)}\over
F}R(\rho)+F^{-2}{\delta^2
\over \delta\phi^2}+(\phi,_u)^2+{1\over F}[(\phi,_\rho)^2+m^2\phi^2]
\right\}\Psi(F, \phi)=0~. \eqno(27)
$$
here it is understood that the discretization of the variable $u$
is already made in an analogous way to that of the previous section.

To make an analysis of this equation we use the following
decomposition
in modes of the scalar field motivated by the solution of the
Klein - Gordon equation, i.e. the semiclassical regime
$$
\phi_k=\phi_k^0\exp\{-ik_+\int_{-\infty}^u
F(\rho,\tilde u)d\tilde
u~\}~~;~~\phi_k^0=e^{-ik_+v}e^{ik_-u}e^{ik\rho}~,
$$
where $-k_+k_-+k^2=-m^2$.

Thus, replacing $\phi_{,u}$ and $\phi_{,\rho}$ into \Eq{27} we can
write approximately (up to a negligible logarithmic term proportional
to $\ln\mid\phi_k/\phi_k^0\mid$),
$$
\left\{-{\delta^2\over\delta F^2}+F^{-2}{\delta^2
\over \delta\phi^2}-U_k(F,\phi)\right\}\Psi(F, \phi)=0~. \eqno(34)
$$
where the effective potential $U_k(F,\phi)$ is given by
$$
U_k(F,\phi)={2~^{(3)}R(\rho)\over F}+\left[k_-^2-2k_+k_-F+k_+^2F^2+
(k^2-m^2)F^{-1}\right]\phi^2~,\eqno(35)
$$
If we use the following transformation of variables
$$
X=F\sinh(\phi)~~,~~T=F\cosh(\phi)~,
$$
this equation can be interpreted as a Klein - Gordon equation for
our wave function with a time - dependent potential $U_k(X,T)$.
When  $U_k<0$ the wave function $\Psi$ will have a {\it real}
exponential
behavior, this corresponds to a classically forbidden regime, in
which
geometries have Euclidean signature. For  $U_k>0$, instead, the wave
function has an exponentially oscillatory behavior corresponding to a
Lorentzian signature (classically allowed) regime.

The analysis of the potential can be done directly in terms of the
variables $F$ and $\phi$. To this end we define:
$$
\alpha\doteq-{k_+
\over\phi^2k_-^3}\left[2~^{(3)}R+(k^2-m^2)\right]\phi^2~~,
{}~~\beta\doteq{k_+F\over k_-}~.
$$
Thus,
$$
U_k=k_-^2\phi^2\left[{-\alpha\over\beta}+(1-\beta)^2\right]~.\eqno(uk
)
$$

The zeros of the potential can thus be obtained from the condition
$$
\alpha_0=\beta_0(1-\beta_0)^2~,\eqno(42)
$$
and the extrema of the potential from
$$
\alpha_m=2\beta_m^2(1-\beta_m)~.\eqno(43)
$$

Let us consider now the potential as a function of $F$ for slices of
$\phi=constant$. There are two main cases:

\noindent
{\bf a) $\alpha<0$}

In this case the potential $U_k$ will always be bigger than zero
having
a minimum at a value $\beta_m>1$ (solving the third order equation
\(43)).
 Thus, we have here the normal oscillatory behavior of the wave
function.

\noindent
{\bf b) $\alpha>0$}

Here we have three subcases depending on the value of $\alpha$:

i) $\alpha>8/27$: The potential grows monotony from $-\infty$ at
$\beta=0$ to $+\infty$ at $\beta\to\infty$ passing through zero
at $\beta_0>1$.

ii) $4/27<\alpha<8/27$: In this case the potential shows a local
maximum
and minimum, both in the region $U_k<0$. For $\beta>\beta_0>1$,
$U_k>0$,
and we have a steady growing. The maximum and minimum are located at
$0<\beta_{max}<2/3$ and $2/3<\beta_{min}<1$ respectively, and they
can be
obtained by from \Eq{43}.

iii) $0<\alpha<4/27$: This case presents two regions where $U_k>0$.
These
two regions are $\beta_0^1<\beta<\beta_0^2$, where $0<\beta_0^1<1/3$
and $1/3<\beta_0^2<1$; and $\beta>\beta_0^3>1$. The two local
maxima and minima are of course located at $\beta_0^1<\beta_{max}
<\beta_0^2$ and $\beta_0^2<\beta_{min}<\beta_0^3$.

\bigskip

The dependence of $U_k$ on $\phi$ is clearly a growing parabola (see
\Eq{35}).
The minimum value is reached at the origin $\phi=0$ and it will
eventually
cross a zero of the potential iff $~^{(3)}R<0$.

Thus, we have seen that we can classify the two regimes: Euclidean
(exponential behavior of the wave function) and Lorentzian
(oscillatory
behavior) according to the sign of the effective potential $U_k$. A
weak potential $U_k$ corresponds to the semiclassical regime. This is
precisely what happens when $F$ is large, i.e. $S\to\infty$, namely
a strong curvature regime, for instance a state picked around a
singular gravitational wave. A strong potential $U_k$ corresponds
to a truly quantum regime in which $S\to0$, i.e. $F\to0$, that is, a
smooth and weak curvature state.

If we prepare the system in a classical configuration, for instance
$|\phi |=1$ and $F\to\infty$ (that would correspond to the
semiclassical
problem studied in \Ref{dv,np1} of a Klein - Gordon field $\phi$ in a
classical shock wave geometry), we observe that we fall into the
case a), i.e.  $U_k>0$, since $~^{(3)}R>0$. Due to the particle
creation (that we
will consider in the next section) of the field $\phi$, the system
evolves towards the minimum of the potential $U_k$ making $F$ small
and $|\phi|$ big, that is, the system evolves towards a quantum
regime
of smooth and weak curvature.

\head{4. Third Quantization}

In this section we study the possibility of third quantizing the
wave function of the ultrarelativistic particles. A motivation
for the third quantization is to overcome the problem
of negative probabilities just as was
the case for the Klein - Gordon equation (see for example \ref{MG}).

Let us then consider the \Eq{27} for an ultrarelativistic
particle and an homogeneous massless scalar field $\phi$, that up to
the operator ordering ambiguity reads,
$$
\left[\partial^2_t-\partial^2_\phi+72~^{(3)}R\exp\{-6t\}\right]
\Psi(t,\phi)=0~,\eqno(49)
$$
where we have made the change of variables (at each point $u_n$)
$$
t_n=-{1\over6}\ln(F_n)~,\eqno(48)
$$
which allows the variable $t_n$ to run from $-\infty$ to $+\infty$
and
to give to it a time - like interpretation.

The general solution to \Eq{49} can be written as
$$
\Psi_{j}(t,\phi)=\prod_nZ_{2ij}\left(\sqrt{8~^{(3)}R}e^{3t_n}\right)
\exp\{ij\phi\}~,\eqno(47)
$$
where $Z_\nu$ is a Bessel function of first or second kind.

It is interesting to note here that the case of pure gravity is
recovered
for $j=0$. As $j$ is related to the energy of the matter field
$\phi$,
it seems that the presence of matter fields excite the gravitational
modes and allow for particle production as we shall immediately see.

Following \ref{ponjas} we can define $in$ states proportional to
the Bessel function $J_{2ij}\left(\sqrt{8~^{(3)}R}e^{3t_n}\right)$
for $t\to-\infty$ (which are natural positive frequency modes
in the $in$ region), and $out$ states
proportional to the Hankel
function $H^{(2)}_{2ij}\left(\sqrt{8~^{(3)}R}e^{3t_n}\right)$
for $t\to+\infty$. Then, one can compute
the Bogoliubov transformation coefficients between these two basis,
and since the Hankel function is a linear combination of positive and
negative subindex (frequency) Bessel functions one
obtains particle production of the outgoing modes with respect to the
$in$
vacuum. It appears that the spectrum of produced particles has
a Planckian distribution
at temperature, $T=18\sqrt{8\pi G/3}$, of the order of
the Planck temperature.

Thus, the interpretation of our results can be made in terms of the
creation of ultrarelativistic particles carrying with them its own
geometry. The interaction with the matter fields deplete the
gravitational
energy bringing the ingoing semiclassical state, for instance picked
around a
shock wave metric, (that is, a strongly curved geometry) into
a state in the quantum regime represented by a weakly curved geometry
(essentially flat spacetime). The evolution of the system is from the
classical into the quantum regime. The initial semiclassical
configuration is
a highly curved geometry, the final configuration is a weakly curved
one.

\bigskip
\bigskip
\noindent
ACKNOWLEDGMENTS

\noindent
The authors thank the Ettore Majorana Center at Erice and IAFE at
Buenos
Aires, where part of this
work was done, for kind hospitality and working facilities. \mangos

\vfill\eject
\references
\baselineskip=18pt

\refis{1} G. 't Hooft,\plb 198,61,1987.

\refis{2} D. Amati, M. Ciafaloni and G. Veneziano,\ijmpa 3,1615,1988.

\refis{6} C. O. Lousto and N. S\'anchez,\ijmpa 5,915,1990.

\refis{8} C. O. Lousto and N. S\'anchez,\prd 46,4520,1992.

\refis{9} C. O. Lousto and N. S\'anchez,\npb 383,377,1992.

\refis{dv} H. J. de Vega and N. S\'anchez,\npb 317,706 and 731,1989.

\refis{dvs} H. J. de Vega and N. S\'anchez,\plb 244,215,1990.

\refis{7} G. T. Horowitz and A. R. Steif,\prl 64,260,1990.

\refis{4} C. O. Lousto and N. S\'anchez,\plb 220,55,1989.

\refis{5} C. O. Lousto and N. S\'anchez,\plb 232,462,1989.

\refis{np1} C. O. Lousto and N. S\'anchez,\npb 355,231,1991.

\refis{ML} C. O. Lousto and F. D. Mazzitelli,\ijmpa 6,1017,1991.

\refis{MG} M. Mc Guigan,\prd 38,3031,1988. ;
 M. Mc Guigan,\prd 39,2229,1989.

\refis{ponjas} A. Hosoya and M. Morikawa,\prd 39,1123,1989.

\refis{3} D. Amati and  C.Klim\v cik,\plb 210,92,1988.

\refis{wdw} B. S. De Witt,\pr 160,1113,1967. ;
J. A. Wheeler, in {\it Batelles Rencontres}, ed. C. De Witt and  J.
A.
Wheeler, Benjamin, N.Y. (1968)\par

\endreferences
\end